\documentstyle[11pt,newpasp,twoside,epsf]{article}
\begin{document}

\title{A Deep Keck Spectral Survey of Cl0024+1654 at ${\bf z=0.4}$}

\author{Anne J. Metevier and David C. Koo}
\affil{Department of Astronomy and Astrophysics, University of California,
 Santa Cruz, CA 95064 USA}

\author{Luc Simard}
\affil{Steward Observatory, University of Arizona,
Tucson, AZ, 85721 USA}

\author{Gregory D. Wirth}
\affil{W. M. Keck Observatory,
Kamuela, HI 96743 USA}

\begin{abstract}
	We have undertaken a Keck spectral survey of galaxies to R$\sim$24
near the central region 
of Cl0024+1654 at $z=0.4$.  The spectral
resolution and slits have been selected to yield kinematics such as the 
velocity dispersions of early-type systems and rotation curves of disks.  The 
spectra also provide a suite of lines that yield 
star formation, age, and metallicity information.
\end{abstract}

\section{Introduction}

	Cl0024 is a rich, intermediate-redshift cluster ideal for galaxy 
evolution studies.  One of the two clusters studied by Butcher and Oemler 
(1978), Cl0024 was noted for its high fraction of ``blue" galaxies relative to 
local clusters.  We have surveyed the central region of this cluster
with the LRIS spectrograph to examine the
internal kinematics of cluster members as well as the emission signatures
of blue disk systems.

\section{LRIS Spectra}

	We have acquired Keck spectra of over 270 objects in the Cl0024 field.
In Table 1, we compare our data to other recent spectroscopic 
surveys of this cluster.  Note that our spectral wavelength range covers
redshifted [OII] through H$\alpha$, and our slits are aligned with
disk major axes.  We have determined redshifts of nearly 150 cluster members
and have confirmed the redshift of the lens source behind this cluster 
at $z=1.68$ (Broadhurst et al. 2000). 
Ellis, Treu, and collaborators are currently undertaking a spectral survey 
of galaxies in the outer regions of Cl0024.  Our data will nicely 
complement theirs as the majority of the galaxies we have surveyed are 
concentrated in the cluster center.  

%\begin{figure}
%\plotone{/net/babar/j/anne/cl0024/lris97/n1/earlytype_figure.ps}
%\caption{Example early- and late-type spectra.}
%\end{figure}

\begin{table}
\caption{Our survey compared to recent spectral surveys of Cl0024.}
\begin{tabular}{lccc}
\tableline
 &              Dressler et al. 1999 &  Czoske et al. 2001 &    {\bf Our Survey}\\
\tableline
        Survey area &   $9.4\arcmin\times7.9\arcmin$    & $21\arcmin\times25\arcmin$    &       $6.1\arcmin\times10.3\arcmin$\\
        N$_{tot}$ &        130          &       687     &       278\\
        N$_{cluster}$ &    107           &       295     &       145\\
        depth   &         $r < 22$      &       $V < 23$ &      ${\bf R < 24.5}$\\
        range (\AA) &   $3500-9800$     &       $4500-8500$ &   $4500-9500$\\
        resolution($\arcsec$/pix) & $0.27-0.59$      &       $0.31-0.44$ &   ${\bf 0.215}$\\
        resolution(\AA/pix) & $2.3-5.3$     &       $3.3-5.0$       &       ${\bf 1.2-1.85}$\\
\tableline
\tableline
\end{tabular}
\end{table}

\section{Efforts Underway}

\indent \indent	{\bf Tully-Fisher study}: We have measured the rotation velocities of seven 
Cl0024 members in a pilot study to characterize the Tully Fisher relation (TFR)
for intermediate-$z$ cluster galaxies (see Metevier et al., also in these
proceedings).  Our data show no evidence for 
evolution of the slope or zeropoint of the cluster TFR.   However, they
do indicate that there may have been more scatter in the past.  We 
intend to strengthen our study with measurements from 15 more cluster spirals
which exhibit spatially resolved emission.  \\

%A robust measurements of the 
%cluster TFR at intermediate redshift will be a powerfl diagnostic of the
%physical processes which affect the evolution of cluster galaxies.

	{\bf Disappearing S0s?}: Dressler et al. (1997, MORPHs), in a study of the
visually classified morphologies of cluster galaxies, have found evidence that the fraction of S0s in clusters has 
doubled since $z=0.5$.  However, other studies of distant cluster S0s suggest that they have red stellar populations similar to those of local cluster galaxies whose stellar populations formed before $z=2$ (Andreon, Davoust, \& Heim 1997).  We have run a preliminary test of the MORPHs
classification of galaxies in the Cl0024 field by measuring the 
bulge-to-total flux ratios (B/T) of these galaxies via a two-component model (deVaucouleurs
bulge + exponential disk).  We find that objects classified by MORPHs
as late types have low B/T, while objects classified as early types tend to 
span a wider range of B/T.  This result suggests that in distant 
clusters, disk-dominated systems are being 
visually classified as bulge-dominated objects (Es and S0s), possibly driving
up counts of S0s in distant clusters.  We are in the process 
of running simulations to test the
reliability of our models.

\end{document}